\documentclass[showpacs,preprintnumbers,amsmath,amssymb,pra,12pt]{revtex4}
\usepackage{graphicx}
\begin{document}

\title{Gaussian modeling and Schmidt modes of SPDS biphoton states}

\author{M V Fedorov$^{1\,*}$, Yu M Mikhailova$^{1,\,2}$ and P A Volkov$^1$}

\address{$^1$A.M.~Prokhorov General Physics Institute,
 Russian Academy of Sciences, 38 Vavilov st., Moscow, 119991, Russia\\
 $^2$Max-Planck-Institut f{\" u}r Quantenoptik, Hans-Kopfermann-Strasse 1,
D-85748, Garching, Germany\\
$^*$E-mail: fedorov@ran.gpi.ru}
\begin{abstract}
A double-Gaussian model and the Schmidt modes are found for the biphoton wave function characterizing spontaneous parametric down-conversion with the degenerate collinear phase-matching of the type I and with a pulsed pump. The obtained results are valid for all durations of the pump pulses, short, long and intermediately long.
\end{abstract}


\maketitle

\section{Introduction}
Biphoton states generated in Spontaneous Parametric Down-Conversion (SPDC) can be highly entangled in continuous variables characterizing signal and idler photons, such as angular variables or frequencies \cite{Angular,Spectral}. For SPDC processes with the type-I phase matching the degree of entanglement is characterized by values of the Schmidt number $K$ up to 300 in the case of short pump pulses. Such high entanglement reflects a high degree of correlations accumulated in the biphoton state and, in principle such highly-correlated states can be useful for goals of quantum information and quantum cryptography. But, of course, there is a big gap between finding that the degree of entanglement is very high and its application in practice. We think that an important step in the direction of closing this gap is finding the Schmidt modes and eigenvalues of the reduced density matrix rather than only the Schmidt number $K$. This task is not quite trivial because Schmidt modes are known only for double-Gaussian wave functions \cite{Wolf,URen}, whereas in some cases even a possibility of modeling the biphoton wave function by a double-Gaussian function is problematic. We will show below that in such cases a reasonable way of derivation consists in calculating first the reduced density matrix (even if approximately) and, then, modeling the latter by a double-Gaussian function. Eigenfunctions of such a model double-Gaussian density matrix are just the Schmidt modes. By continuing this procedure, and using general connections between the density matrix and wave function, we show that the double-Gaussian model of the spectral wave function exists, and it's found explicitly.

The paper is structured in the following way. In the next Section we give a brief overview of the relation between the double-Gaussian wave functions and density matrices and derivation of their Schmidt modes. In Section 3 we summarize the main features of the spectral biphoton wave function for the degenerate collinear SPDC process with a pulsed pump. In Section 4 we describe how the Gaussian modeling can be applied for finding Schmidt modes in the case of long pump pulses. In Section 5 we describe Gaussian modeling of the reduced density matrix and derivation of the Schmidt modes in the case of short pump pulses. In Section 6 we formulate some generalizations of the derived specific results and suggest a single double-Gaussian model of the biphoton spectral wave function valid for arbitrary pump-pulse durations.

\section{Reduced density matrix and Schmidt modes for double-Gaussian bipartite wave functions}
Let $x_1$ and $x_2$ be some continuous variables of particles ``1" and ``2"  and $\Psi(x_1,\,x_2)$ be a bipartite wave function. Lets us consider here only the case of symmetric bipartite wave functions for which  $\Psi(x_1,\,x_2)=\Psi(x_2,\,x_1)$. The most general form of the double-Gaussian wave function satisfying this condition is given by
\begin{equation}
 \label{DG-ab}
 \Psi(x_1,\,x_2)=\sqrt{\frac{2}{\pi a b}}\,\exp\left[-\frac{(x_1+x_2)^2}{2a^2}\right]\, \exp\left[-\frac{(x_1-x_2)^2}{2b^2}\right],
\end{equation}
where $a$ and $b$ are parameters related to widths of distributions in $x_{1,\,2}$ determined by each Gaussian factor in Equation (\ref{DG-ab}), $a>0$ and $b>0$. Alternatively, the same wave function can be presented in a different form, convenient for determining the double-Gaussian Schmidt modes \cite{URen}
\begin{equation}
 \label{DG-mu-alpha}
 \Psi(x_1,\,x_2)=\frac{\alpha}{\sqrt{\pi}}\,\exp\left\{\alpha^2\left[-\frac{1+\mu^2}{2(1-\mu^2)}(x_1^2+x_2^2)
 +\frac{2\mu}{(1-\mu^2)}x_1x_2\right]\right\}.
\end{equation}
The parameters $a$ and $b$ are related to $\mu$ and $\alpha$ by equations
\begin{equation}
 \label{ab-mu,alpha}
 \mu=\frac{a-b}{a+b},\,\alpha=\sqrt{\frac{2}{ab}}\quad{\rm and}\quad a=\frac{\sqrt{2}}{\alpha}\sqrt{\frac{1+\mu}{1-\mu}}\,,\,b=\frac{\sqrt{2}}{\alpha}\sqrt{\frac{1-\mu}{1+\mu}}\,,
\end{equation}
$-1<\mu<1$.

The density matrix and reduced density matrix, corresponding to the wave function $\Psi(x_1,\,x_2)$, are given by
\begin{eqnarray}
 \nonumber
 \rho(x_1,\,x_2;\,x_1^\prime,\,x_2^\prime)=\Psi(x_1,\,x_2)\Psi(x_1^\prime,\,x_2^\prime)\quad{\rm and}\\
 \label{DM}
 \rho_r(x_1,\,x_1^\prime)=\int dx_2\rho(x_1,\,x_2;\,x_1^\prime,\,x_2).
\end{eqnarray}
After integration, the reduced density matrix can be presented in the same forms as described above for the wave function, (\ref{DG-ab}) and (\ref{DG-mu-alpha})
\begin{eqnarray}
 \nonumber
 \rho_r(x_1,\,x_1^\prime)=\frac{1}{{\widetilde a}}\,\sqrt{\frac{2}{\pi}}\,\exp\left[-\frac{(x_1+x_1^\prime)^2}{2{\widetilde a}^2}\right]\, \exp\left[-\frac{(x_1-x_1^\prime)^2}{2{\widetilde b}^2}\right] \\
 \label{rho-r}
 ={\widetilde \alpha}\sqrt{\frac{2}{\pi}\,\frac{1+{\widetilde \mu}}{1-{\widetilde \mu}}}\,\exp\left\{{\widetilde \alpha}^2\left[-\frac{1+{\widetilde \mu}^2}{2(1-{\widetilde \mu}^2)}(x_1^2+{x_1^\prime}^2)+\frac{{2\widetilde \mu}}{(1-{\widetilde \mu}^2)}x_1 x_1^\prime\right]\right\}.
\end{eqnarray}
The pairs of parameters ${\widetilde a},{\widetilde b}$ and ${\widetilde \alpha},{\widetilde \mu}$ are related to each other by the same formulas as
$a,b$ and $\alpha,\mu$ (\ref{ab-mu,alpha}):
\begin{equation}
 \label{tilde-ab-mu,alpha}
 {\widetilde\mu}=\frac{{\widetilde a}-{\widetilde b}}{{\widetilde a}+{\widetilde b}},\,{\widetilde \alpha}=\sqrt{\frac{2}{{\widetilde a}{\widetilde b}}}\quad{\rm and}\quad {\widetilde a}=\frac{\sqrt{2}}{{\widetilde \alpha}}\sqrt{\frac{1+{\widetilde \mu}}{1-{\widetilde \mu}}},\,{\widetilde b}=\frac{\sqrt{2}}{{\widetilde \alpha}}\sqrt{\frac{1-{\widetilde \mu}}{1+{\widetilde \mu}}}\,.
\end{equation}
Moreover, with a simple algebra we find a series of formulas relating the parameters characterizing the density matrix with the parameters characterizing the wave function
\begin{equation}
 \label{DM-WF param}
 {\widetilde \alpha}=\alpha,\, {\widetilde\mu}=\mu^2,\,{\widetilde a}=\sqrt{\frac{a^2+b^2}{2}},\, {\widetilde b}=ab\sqrt{\frac{2}{a^2+b^2}},
\end{equation}
and
\begin{equation}
 \label{DM-WF param-2}
 \left\{{a=\sqrt{\displaystyle\frac{{\widetilde a}}{2}}\left(\sqrt{{\widetilde a}+{\widetilde b}}-\sqrt{{\widetilde a}-{\widetilde b}}\,\right)}
 \atop{b=\sqrt{\displaystyle\frac{{\widetilde a}}{2}}\left(\sqrt{{\widetilde a}+{\widetilde b}}+\sqrt{{\widetilde a}-{\widetilde b}}\,\right).}\right.
\end{equation}
The Schmidt modes $\{\psi_n(x_1),\chi_n(x_2)\}$ are defined as eigenfunctions of the reduced density matrices $\rho_r(x_1,\,x_1^\prime)$ and
$\rho_r(x_2,\,x_2^\prime)$. For symmetric wave functions $\Psi(x_1,\,x_2)$ two functions in the Schmidt-mode pairs are identical, $\chi_n(x_2)=\psi_n(x_2)$. In a general case, to find Schmidt modes one has to solve the integral equation
\begin{equation}
 \label{int-eq}
 \int dx_1^\prime \rho_r(x_1,\,x_1^\prime)\,\psi_n(x_1^\prime)=\lambda_n\,\psi_n(x_1),
\end{equation}
where $\lambda_n$ are eigenvalues of the reduced density matrix. The Schmidt modes diagonalize the reduced density matrix and, hence, the latter can be presented as the sum of products of Schmidt modes
\begin{equation}
 \label{Diagonal RDM}
 \rho_r(x_1,\,x_1^\prime)=\sum_{n=0}^\infty \lambda_n\,\psi_n(x_1)\psi_n(x_1^\prime).
\end{equation}
According to the Schmidt theorem, the wave function $\Psi(x_1,\,x_2)$ also can be expanded in a series of products of the Schmidt modes
\begin{equation}
 \label{Schmidt theorem}
 \Psi(x_1,\,x_2)=\sum_{n=0}^\infty \sqrt{\lambda_n}\;\psi_n(x_1)\psi_n(x_2).
\end{equation}
For the double-Gaussian wave functions analytical expressions for the Schmidt modes are known and they follow from the formula \cite{URen}
\begin{gather}
 \nonumber
 \exp\left\{\alpha^2\left[-\frac{1+\mu^2}{2(1-\mu^2)}(x_1+x_2)^2+\frac{2\mu}{(1-\mu^2)}x_1x_2\right]\right\}\\
 \label{Exp-decompos}
 =\sqrt{\pi}\,\sqrt{1-\mu^2}\sum_{n=0}^\infty \mu^n \varphi_n(\alpha x_1)\varphi_n(\alpha x_2),
\end{gather}
where $\varphi_n$ are the oscillator wave functions $\varphi_n(x)=(2^n n!\sqrt{\pi})^{-1/2}e^{-x^2/2}H_n(x)$, $H_n(x)$ are the Hermite polynomials. The Schmidt modes and eigenvalues of the reduced density matrix are easily found now from Equations (\ref{DG-ab}), (\ref{DG-mu-alpha}), (\ref{ab-mu,alpha}), and (\ref{Exp-decompos})
\begin{gather}
 \nonumber
 \psi_n(x_{1,\,2})=\sqrt{\alpha}\,\varphi_n(\alpha x_{1,\,2})\\
 \label{eigenmodes}
 =\left(\frac{2}{ab}\right)^{1/4}\,\varphi_n\left(\frac{\sqrt{2}\,x_{1,\,2}}{\sqrt{ab}}\right)
  =\left(\frac{2}{{\widetilde a}{\widetilde b}}\right)^{1/4}\,\varphi_n\left(\frac{\sqrt{2}\,x_{1,\,2}}{\sqrt{{\widetilde a}{\widetilde b}}}\right)
\end{gather}
and
\begin{gather}
 \nonumber
 \sqrt{\lambda_n}=\sqrt{1-\mu^2}\,\mu^n=\frac{2\sqrt{ab}}{a+b}\,\left(\frac{a-b}{a+b}\right)^n
 =(-1)^n\sqrt{\frac{2{\widetilde b}}{{\widetilde a}+{\widetilde b}}}\left(\frac{{\widetilde a}-{\widetilde b}}{{\widetilde a}+{\widetilde b}}\right)^{n/2},\\
 \label{eigenvalues}
 \lambda_n=4ab\,\frac{(a-b)^{2n}}{(a+b)^{2(n+1)}}=\frac{2{\widetilde b}}{{\widetilde a}+{\widetilde b}}
 \left(\frac{{\widetilde a}-{\widetilde b}}{{\widetilde a}+{\widetilde b}}\right)^n.
\end{gather}
The Schmidt number, characterizing the degree of entanglement in the state $\Psi(x_1,\,x_2)$ is defined as
\begin{equation}
 \label{K-DG}
 K=\left(\sum_{n=0}^\infty \lambda_n^2\right)^{-1}=\frac{1+\mu^2}{1-\mu^2}=\frac{1+{\widetilde\mu}}{1-{\widetilde\mu}}
 =\frac{a^2+b^2}{2ab}=\frac{\widetilde a}{\widetilde b}\,,
\end{equation}
which agrees with more general expressions of Refs. \cite{URen,2006}.

\section{Spectral biphoton wave function}
Let us remind now the main features of the biphoton spectral wave function for SPDC with the type-I degenerate collinear phase matching and with a pulsed pump.
The main general expression has the form \cite{Spectral,Rubin}
\begin{gather}
 \nonumber
 \Psi(\nu_1,\,\nu_2)\propto \exp\left(-\frac{(\nu_1+\nu_2)^2\tau^2}{8\ln
 2}\right)\\
 \times
 {\rm  sinc}\left\{\frac{L}{2c}\left[A(\nu_1+\nu_2)-B\frac{(\nu_1-\nu_2)^2}{\omega_0}\right]
 \right\},\label{WF}
\end{gather}
where  $\tau$ is the pump-pulse duration, $L$ is the length of a crystal, $\nu_1$ and $\nu_2$ are deviations of frequencies of the signal and idler photons $\omega_{1,\,2}$ from the central frequencies $\omega_1^{(0)}=\omega_2^{(0)}=\omega_0/2$, $|\nu_{1,\,2}|\ll\omega_0$, $\omega_0$ is the central frequency of the pump spectrum, $A$ and $B$ are the temporal walk-off and dispersion constants
\begin{gather}
 \nonumber
 A=c\left(\left.k_p^\prime(\omega)\right|_{\omega=\omega_0}-
 \left.k_1^\prime(\omega)\right|_{\omega=\omega_0/2}\right)=c\left(\frac{1}{{\rm v}_g^{(p)}}-\frac{1}{{\rm v}_g^{(o)}}\right)\label{A},\\
 \label{AB}
 B=\frac{c}{4}\,\omega_0\left.k_1^{\prime\prime}(\omega)\right|_{\omega=\omega_0/2},
\end{gather}
${\rm v}_g^{(p)}$ and ${\rm v}_g^{(o)}$ are the group velocities
of the pump and ordinary waves, and $k_1$ and $k_p$ are the wave vectors of signal and pump photons.

The first term on the right-hand side of Equation (\ref{WF}) is the pump spectral field strength taken in the Gaussian form. The second term is the sinc-function characterizing the crystal and propagation in it of the pump and signle/idler photons. The argument of the sinc-function contains both linear and quadratic terms in $\nu_{1,\,2}$. As we assume that $|\nu_{1,\,2}|\ll\omega_0$, in principle, the quadratic term is much smaller than the linear one. Nevertheless, the quadratic term can never be dropped because it is crucially important for determining the finite-width single particle spectrum [without the quadratic term both factors on the right-hand side of Equation (\ref{WF}) depend only on $\nu_1+\nu_2$, which makes the integral $\int d\nu_2|\Psi(\nu_1,\,\nu_2)|^2$ independent of $\nu_1$ and single-particle spectrum infinitely wide]. As for the linear term in the argument of the sinc-function it can be efficiently eliminated in the case of long pump pulses but cannot be dropped if pulses are short. The control parameter separating the regions of short and long pulses is given by \cite{Angular}
\begin{equation}
 \label{eta}
 \eta=\frac{\Delta\nu_{1\,{\rm sinc}}}{\Delta\nu_{1\,{\rm pump}}}
 \approx 2\frac{c\,\tau}{AL}=\frac{2\tau}{L/{\rm v}_g^{(p)}-L/{\rm v}_g^{(o)}},
\end{equation}
i.e., it's equal to the ratio of the double pump-pulse duration to the difference of times required for the the pump and idler/signal photons for traversing all the crystal. Pump pulses are short if $\eta\ll 1$ and long if $\eta\gg 1$ and, typically, $\eta\sim 1$ at $\tau\sim 1 {\rm ps}$.

The degree of entanglement of the state (\ref{WF}) was characterized by two parameters, the Schmidt number $K$ and the parameter $R$ defined as the ratio of the single- to coincidence spectral widths of the corresponding photon distributions, $R=\Delta\nu_1^{(s)}/\Delta\nu_1^{(c)}$ \cite{Spectral}.
\begin{figure}
[h]
\centering\includegraphics[width=6cm]{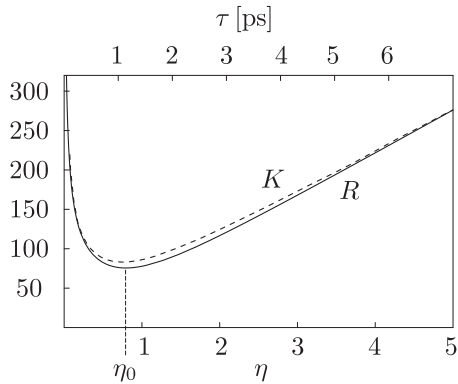}
\caption{{\protect\footnotesize {Entanglement parameters: $R$, calculated analytically \cite{Spectral}, and $K$, calculated numerically \cite{Mauerer}, for a crystal LiIO$_3$ of the length $L=0.5\,{\rm cm}$ and the type-I phase matching.}}}\label{Fig1}
\end{figure}
Both parameters were calculated analytically as function of the pump-pulse duration $\tau$ and were found to be rather close to each other. Also, for the same wave function (\ref{WF}), the Schmidt number $K(\tau)$  was calculated numerically \cite{Mauerer}, and the result was found to be in a very good agreement with the analytical function $R(\tau)$ of Ref. \cite{Spectral} (see Fig. \ref{Fig1}). The degree of spectral entanglement was found to be rather high at all values of the pump-pulse duration and especially high in the cases of sufficiently short and long pulses (far from the minimum which was found to occur at $\eta\sim 1$ or $\tau\sim 1\,{\rm ps}$). Schmidt modes were not calculated for the wave function of the form (\ref{WF}), either analytically or numerically. Below we present an approximate analytical derivation of the Schmidt modes separately in the cases of long and short pump pulses.

For further analysis we will need explicit analytical expressions for the coincidence and single-particle spectral widths of the SPDC biphoton state (\ref{WF}) in the limit of short pump pulses, $\eta\ll 1$. They were found in the work \cite{Spectral} to be given by
\begin{equation}
 \label{width-short}
 \Delta\omega^{(c)}_{\rm shotrt}=\frac{5.56\,c}{AL}, \quad\Delta\omega^{(s)}_{\rm shotrt}=\sqrt{\frac{2A\ln (2)\,\omega_0}{B\tau}}\,.
\end{equation}

Note finally that a very small difference between the Schmidt number $K(\tau)$ (exact and numerical,  \cite{Mauerer})  and the parameter $R(\tau)$ (analytical, \cite{Spectral}) raises often a question why are they so close to each other whereas the wave function (\ref{WF}) is rather significantly non-double-Gaussian? In this work we give an answer: actually, the wave function (\ref{WF}) is ``hiddenly double-Gaussian$"$. And we make this hidden double-Gaussian character of the wave function (\ref{WF}) explicit by finding the double-Gaussian models for the regions of short and long pump pulses and by generalizing these findings for all values of the pump pulse duration (Section 6).

\section{Gaussian modeling and Schmidt modes in the case of long pump pulses}
In the case of long pump pulses, $\eta\gg 1$, the Gaussian function on the right-hand side of Equation (\ref{WF}) is much narrower the sinc-function. For this reason $\nu_1\approx -\nu_2$, and this eliminates efficiently the term linear in $\nu_1$ and $\nu_2$ in the argument of the sinc-function and reduces $\Psi(\nu_1,\,\nu_2)$ to the form
\begin{equation}
 \label{WF-long-appr}
 \Psi(\nu_1,\,\nu_2)\propto \exp\left[-\frac{(\nu_1+\nu_2)^2\tau^2}{8\ln
 2}\right]\,{\rm  sinc}\left[\frac{BL}{2c\omega_0}\,(\nu_1-\nu_2)^2\right].
\end{equation}
The wave function of a similar form was considered earlier in the analysis of angular entanglement of SPDC biphoton states \cite{Law-Eberly}, and then some first Schmidt modes were found numerically. On the other hand, the form of the sinc-function in Equation (\ref{WF-long-appr}) is already very convenient for Gaussian modeling. Parameters of the corresponding Gaussian function can be chosen from the coincidence condition of the Full Widths at Half Maxima (FWHM) of the squared functions sinc($u^2$) and $\exp(-\gamma u^2)$ which gives $\gamma=0.249$. Thus, the substitution
\begin{equation}
 \label{sin-Gauss}
 {\rm  sinc}\left[\frac{BL}{2c\omega_0}\,(\nu_1-\nu_2)^2\right]\rightarrow
 \exp\left[-0.249\,\frac{B\,L}{2c\,\omega_0}\,(\nu_1-\nu_2)^2\right],
\end{equation}
reduces the wave function of Equation (\ref{WF-long-appr}) to the double-Gaussian form (\ref{DG-ab})
\begin{equation}
 \Psi(\nu_1,\,\nu_2)= \sqrt{\frac{2}{\pi ab}}\exp\left[-\frac{(\nu_1+\nu_2)^2\tau^2}{2a^2}\right]
 \label{wf-long-G}
 \exp\left[-\frac{(\nu_1-\nu_2)^2}{2b^2}\right],
\end{equation}
with the parameters $a$ and $b$ given by
\begin{equation}
 \label{ab-long}
 a=\frac{2\sqrt{\ln 2}}{\tau}\quad {\rm and}\quad b=\sqrt{\frac{c\,\omega_0}{0.249\,BL}}=c\,\sqrt{\frac{2\pi}{0.249\,BL\lambda_0}}\;,
\end{equation}
where $\lambda_0$ is the central wavelength of the pump. Note, that owing to the assumption $\eta\gg 1$ and also because always $L\gg\lambda_0$, the ratio of the parameters $b$ and $a$ is large: $b/a\sim \eta\sqrt{L/\lambda_0}\gg 1$. By using Equations (\ref{ab-mu,alpha}) we find also other parameters of the double-Gaussian wave function, $\mu$ and $\alpha$:
\begin{equation}
 \label{mu-long}
 \mu_{\rm long}=-\frac{b-a}{b+a}\approx -1+2\frac{a}{b}=-1+\frac{2}{c\tau}\sqrt{\frac{0.249\ln 2\,BL\lambda_0}{2\pi}}
\end{equation}
and
\begin{equation}
 \label{alpha-long}
 \alpha_{\rm long}=\sqrt{\frac{2}{ab}}=\sqrt\frac{\tau}{c}\left(\frac{0.249\,BL\lambda_0}{2\pi\ln 2}\right)^{1/4}.
\end{equation}
The Schmidt number (\ref{K-DG}) for long pump pulses appears to be given by
\begin{equation}
 \label{K-long}
 K_{\rm long}=\frac{a^2+b^2}{2ab}\approx\frac{b}{2a}=\frac{\sqrt{2\pi}}{4\sqrt{0.249\ln 2}}\;\frac{c\tau}{\sqrt{\lambda_0 L\,B}}\approx 1.5084\frac{c\tau}{\sqrt{\lambda_0 L\,B}}\,.
\end{equation}
This result coincides perfectly with the earlier derived expression for the parameter $R$ in the approximation of long pump pulses \cite{Spectral}
\begin{equation}
 \label{R-long}
 R_{\rm long}=\frac{\sqrt{1.39\,\pi}}{2\ln 2}\;\frac{c\tau}{\sqrt{\lambda_0 L\,B}}\approx 1.50739\frac{c\tau}{\sqrt{\lambda_0 L\,B}}\,.
\end{equation}
Though the coefficients in Equations (\ref{K-long}) and (\ref{R-long}) look different, numerically they are seen to be amazingly close to each other. Note that the parameter $R_{\rm long}$ was found in the paper \cite{Spectral} without any Gaussian modeling and in the frame of a procedure absolutely different from that used above for the derivation of the Schmidt number $K_{\rm long}$.

In terms of $K_{\rm long}$, Equations (\ref{mu-long}) and (\ref{alpha-long}) for the parameters $\mu$ and $\alpha$ can be rewritten as
\begin{equation}
 \label{mu-alpha-long}
 \mu_{\rm long}=\approx -1+\frac{1}{K_{\rm long}}\quad{\rm and}\quad\alpha_{\rm long}\approx\frac{1}{a\sqrt{K_{\rm long}}}
 =\frac{\tau}{2\sqrt{K_{\rm long}\ln 2}}.
\end{equation}

Again in terms of $K_{\rm long}$, the Schmidt modes (\ref{eigenmodes}) and eigenvalues of the reduced density matrix (\ref{eigenvalues}) take the form
\begin{equation}
 \label{eigenmodes-long}
 \psi_n(\nu_{1,\,2})=\frac{\sqrt{\tau/2}}{(K_{\rm long}\ln 2)^{1/4}}
 \varphi_n\left[ \frac{\tau\nu_{1,\,2}}{2\sqrt{K_{\rm long}\ln 2}}\right]
\end{equation}
and
\begin{equation}
\label{eigenvalues-long}
 \sqrt{\lambda_n}\approx (-1)^n\sqrt{\frac{2}{K_{\rm long}}}\left(1-\frac{1}{K_{\rm long}}\right)^n,\quad
 \lambda_n\approx\frac{2}{K_{\rm long}}\left(1-\frac{2}{K_{\rm long}}\right)^n.
\end{equation}

The derivations of this section were based on a direct Gaussian modeling of the wave function, without explicit consideration of the density matrix. As we will see in the following section, in the case of short pump pulses such simple procedure does not work, and to apply Gaussian modeling, one has to calculate first the reduced density matrix. This indicates a rather important qualitative difference between the cases of long and short pump pulses.

\section{Schmidt modes in the case of short pump pulses}
In the case of short pump pulses we cannot use anymore the above-described procedure of a direct Gaussian modeling of the wave function (\ref{WF}). Indeed, in this case the pump spectral function is much wider than the sinc-function and, hence, the linear term in the sinc-function's argument cannot be eliminated. As for the quadratic term, it cannot be dropped too because it is crucially important for appropriate determination of the single-particle spectrum. As for the sinc-function of Equation (\ref{WF}) with its full argument, it's rather difficult to see directly any possibility of its replacement by any Gaussian function. For example, a simple substitution of sinc$(u)$ by $e^{-u^2}$ gives a super-Gaussian function with quadratic and both third- and forth-power terms in $\nu_{1,2}$ under the symbol of exponent. This is not a simplification at all and does not provide a way for analytical derivation of the Schmidt modes. So, we will apply here a somewhat subtler method. At first, we will calculate approximately the reduced density matrix corresponding to the wave function (\ref{WF}). The reduced density matrix found in such a way appears to be much more appropriate for double-Gaussian modeling than the wave function. For the Gaussian model of the reduced density matrix, its eigenfunctions (Schmidt modes), eigenvalues and the Schmidt number are easily found. After this, with the help of equations of Section II, we will find parameters of the effective double-Gaussian model for the original wave function (\ref{WF}), which appears to be very non-trivial and hardly guessable in advance.

\subsection{Density matrix}
In a general form the density matrix corresponding to the wave function (\ref{WF}) is given by
\begin{gather}
 \nonumber
 \rho_r(\nu_1,\,\nu_1^{\,\prime})\propto
 \int d\nu_2 \exp\left\{-\frac{\left[(\nu_1+\nu_2)^2+(\nu_1^{\,\prime}+\nu_2)^2\right]\tau^2}{8\ln
 2}\right\}\\
 \nonumber
 \times {\rm  sinc}\left\{\frac{L}{2c}\left[A(\nu_1+\nu_2)-B\frac{(\nu_1-\nu_2)^2}{\omega_0}\right]\right\}\\
 \label{RDM-short}
 \times{\rm  sinc}\left\{\frac{L}{2c}\left[A(\nu_1^{\,\prime}+\nu_2)-B\frac{(\nu_1^{\,\prime}-\nu_2)^2}{\omega_0}\right]
 \right\}.
\end{gather}
Let us make in this equation a series of approximations similar to those made in Section {\bf V}A of the work \cite{Spectral} and based, actually, on a single physical reason: under the assumption about short duration of pump pulses, $\eta\ll 1$, the Gaussian factor in Equation (\ref{RDM-short}), as a function of either $\nu_1$, $\nu_1^{\,\prime}$, or $\nu_2$, is much smoother than both sinc-functions. Compared to the localization region of the Gaussian function $\sim1/\tau$ both the difference $|\nu_1-\nu_1^{\,\prime}|$ and sums $|\nu_2+\nu_1|$, $|\nu_2+\nu_1^{\,\prime}|$ are very small: they are on the order of $c/AL$, as they are determined by the linear terms in the arguments of the sinc-functions. Owing to this, we can approximate $\nu_2$ by $-\nu_1$ and by $ -\nu_1^{\,\prime}$ in small quadratic terms in the arguments of the first and second sinc-functions in Equation (\ref{RDM-short}), thus linearizing their arguments with respect to $\nu_2$:
\begin{gather}
 \nonumber
 \rho_r(\nu_1,\,\nu_1^{\,\prime})\propto
 \int d\nu_2 \exp\left\{-\frac{\left[(\nu_1+\nu_2)^2+(\nu_1^{\,\prime}+\nu_2)^2\right]\tau^2}{8\ln
 2}\right\}\\
 \times {\rm  sinc}\left\{\frac{AL}{2c}\left[\nu_2-\bar{\nu_2}(\nu_1)\right]\right\}
 {\rm  sinc}\left\{\frac{AL}{2c}\left[\nu_2-\bar{\nu_2}(\nu_1^{\,\prime})\right]\right\},
 \label{RDM-2}
\end{gather}
where
\begin{equation}
 \label{nu2ab}
 \bar{\nu_2}(\nu_1)=-\nu_1+\frac{4B\nu_1^2}{A\omega_0}\quad{\rm and}\quad  \bar{\nu_2}(\nu_1^{\,\prime})=-\nu_1^{\,\prime}+\frac{4B{\nu_1^{\,\prime}}^2}{A\omega_0}.
\end{equation}
The slowly changing exponential function in Equation (\ref{RDM-2}) can be taken out of the integral, for example, at $\nu_2=\frac{1}{2}\left[\bar{\nu_2}(\nu_1)+\bar{\nu_2}(\nu_1^{\,\prime})\right]$ to give
\begin{equation}
 \label{exponent}
 \exp\left\{-\frac{\tau^2}{\ln 2}\left[ \frac{(\nu_1-\nu_1^{\,\prime})^2}{16}+\frac{B^2(\nu_1^2+{\nu_1^{\,\prime}}^2)^2}{A^2\omega_0^2}\right]\right\}
 \approx\exp\left[-\frac{\tau^2B^2(\nu_1+\nu_1^{\,\prime})^4}{4A^2\omega_0^2\ln 2}\right],
\end{equation}
where in the last approximate expression we have dropped all small terms proportional to $(\nu_1-\nu_1^{\,\prime})^2$. In this approximation, instead of $\nu_2=\frac{1}{2}\left[\bar{\nu_2}(\nu_1)+\bar{\nu_2}(\nu_1^{\,\prime})\right]$, we could choose for evaluation of the exponential factor either $\nu_2=\bar{\nu_2}(\nu_1)$ or $\nu_2=\bar{\nu_2}(\nu_1^{\,\prime})$. In all cases the final result would be the same as given by the last expression in Equation (\ref{exponent}).

The remaining integral of the product of two sinc-functions is calculated with the help of Equation (37) of the paper \cite{Spectral}
\begin{gather}
 \nonumber
 \int d\nu_2 \,{\rm  sinc}\left\{\frac{AL}{2c}\left[\nu_2-\bar{\nu_2}(\nu_1)\right]\right\}
 {\rm  sinc}\left\{\frac{AL}{2c}\left[\nu_2-\bar{\nu_2}(\nu_1^{\,\prime})\right]\right\}\\
 \label{sinc-sinc}
 =\pi\,{\rm  sinc}\left\{\frac{AL}{2c}\left[\bar{\nu_2}(\nu_1)-\bar{\nu_2}(\nu_1^{\,\prime})\right]\right\}
 \approx \pi\,{\rm  sinc}\left[\frac{AL}{2c}(\nu_1-\nu_1^{\,\prime})\right].
\end{gather}
By combining Equations (\ref{exponent}) and (\ref{sinc-sinc}) together, we find the following representation for the reduced density matrix
\begin{gather}
  \label{RDM-3}
  \rho_r(\nu_1,\,\nu_1^{\,\prime})\propto \exp\left[-\frac{\tau^2B^2(\nu_1+\nu_1^{\,\prime})^4}{4A^2\omega_0^2\ln 2}\right]\,{\rm  sinc}\left[\frac{AL}{2c}(\nu_1-\nu_1^{\,\prime})\right].
\end{gather}
Note that in the last approximate expression of Equation (\ref{sinc-sinc}) all quadratic terms in the argument of the sinc-function are dropped. Though such approximation could not be used for the wave function (\ref{WF}), it appears to be well applicable for the density matrix (\ref{sinc-sinc}) because the quadratic terms in the argument of the sinc function are small and, most important, because the linear term is proportional to the difference of frequencies $\nu_1-\nu_1^{\,\prime}$ rather than their sum, whereas the exponential factor in Equation (\ref{RDM-3}) depends on $\nu_1+\nu_1^{\,\prime}$. This feature of the density matrix contrasts to that of the wave function (\ref{WF}) in which both the pump spectrum and the linear term in the argument of the sinc-function depend on the sum of frequencies $\nu_1+\nu_2$.

\subsection{Gaussian modeling and Schmidt modes for short pump pulses}
Both factors in the reduced density matrix (\ref{RDM-3}) can be modeled by Gaussian functions by means of substitutions
\begin{equation}
 \label{subsitution-1}
  \exp\left[-\frac{\tau^2B^2(\nu_1+\nu_1^{\,\prime})^4}{4A^2\omega_0^2\ln 2}\right]\rightarrow
  \exp\left[-\gamma_1\,\frac{\tau B(\nu_1+\nu_1^{\,\prime})^2}{2A\omega_0\sqrt{\ln 2}}\right]
\end{equation}
 and
 \begin{equation}
 \label{subsitution-2}
  {\rm  sinc}\left[\frac{AL}{2c}(\nu_1-\nu_1^{\,\prime})\right]\rightarrow
  \exp\left[-\gamma_2\,\frac{A^2L^2(\nu_1-\nu_1^{\,\prime})^2}{4c^2}\right],
\end{equation}
where $\gamma_1$ and $\gamma_2$ are  fitting factors to be defined below.

With these substitutions the reduced density matrix (\ref{RDM-3}) takes the double-Gaussian form (\ref{rho-r})
\begin{equation}
 \label{RDM-G-short}
 \rho_r(\nu_1,\,\nu_1^{\,\prime})=\frac{1}{{\widetilde a}}\,\sqrt{\frac{2}{\pi}}\,\exp\left[-\frac{(\nu_1+\nu_1^{\,\prime})^2}{2{\widetilde a}^2}\right]\, \exp\left[-\frac{(\nu_1-\nu_1^{\,\prime})^2}{2{\widetilde b}^2}\right]
\end{equation}
with
\begin{equation}
 \label{ab-tilde-short}
 {\widetilde a}=(\ln 2)^{1/4}\sqrt{\frac{A\omega_0}{\gamma_1\tau B}}\,,\quad {\widetilde b}=\frac{c}
 {AL}\sqrt{\frac{2}{\gamma_2}}\,.
\end{equation}
In terms of the control parameter $\eta$ (\ref{eta}), roughly,
\begin{equation}
 \label{a-tilde-gg}
 {\widetilde a}\sim{\widetilde b}\,\sqrt{\displaystyle\frac{L}{\lambda_0\eta}}\gg{\widetilde b},
\end{equation}
because for short pulses $\eta\ll 1$ and always $L\gg\lambda_0$.

With the help of Equations (\ref{eigenmodes}), (\ref{eigenvalues}), and (\ref{K-DG}) we can find now other parameters of the double-Gaussian reduced density matrix (${\widetilde\alpha}$ and {$\widetilde\mu$), as well as the Schmidt number $K$. The latter is given by
\begin{equation}
 \label{K-short-gammas}
 K_{\rm short}=\frac{\widetilde a}{\widetilde b}=(\ln 2)^{1/4}\frac{AL}{c}\sqrt{\frac{\gamma_2A\omega_0}{2\gamma_1\tau B}}
 =\frac{A^{3/2}}{\sqrt{B}}\frac{L}{\sqrt{\lambda_0c\tau}}(\ln 2)^{1/4}\sqrt{\frac{\pi\gamma_2}{2\gamma_1}}.
\end{equation}

\subsection{Gaussian modeling of the original wave function}

In accordance with the discussion of Section 2, there is a one-to-one correspondence between the double-Gaussian reduced density matrix and the double-Gaussian wave function. This means that, as we have received the double-Gaussian representation (\ref{RDM-G-short}) for the reduced density matrix of the biphoton state arising in the degenerate collinear SPDC process with the type-I phase matching in the limit of short pump pulses, the original wave function (\ref{WF}) also must be representable in the double-Gaussian form, even though it's not evident in advance. Parameters of the double-Gaussian model of the wave function can be found from Equations (\ref{ab-tilde-short}) and
(\ref{DM-WF param-2}), the latter of which can be simplified in the approximation ${\widetilde a}\gg{\widetilde b}$ to give
\begin{equation}
 \label{ab-short}
 a\approx\frac{\widetilde b}{\sqrt{2}}=\frac{c}{AL\sqrt{\gamma_2}}\,,\;
 b\approx\sqrt{2}\,{\widetilde a}=(\ln 2)^{1/4}\sqrt{\frac{2A\omega_0}{\gamma_1\tau B}}\,,
\end{equation}
with the relation between $a$ and $b$ opposite to that occurring between ${\widetilde a}$ and ${\widetilde b}$ (\ref{a-tilde-gg}):
\begin{equation}
 \label{a-ll-b}
 a\sim b\,\sqrt{\displaystyle\frac{\lambda_0\eta}{L}}\ll b.
\end{equation}
With $a$ and $b$ given by Equations (\ref{ab-short}), the double-Gaussian spectral wave function modeling that of Equation (\ref{WF}) is given by Equation (\ref{DG-ab}) (with $x_{1,2}$ substituted by $\nu_{1,2}$). In the approximation $a\ll b$ (\ref{a-ll-b}) the coincidence and single-particle spectra corresponding to this wave function are determined by the curves
\begin{equation}
 \label{coi-spectrum-short}
 \frac{dw^{(c)}(\nu_1)}{d\nu_1}\propto \exp\left[-\frac{(\nu_1+\nu_2)^2}{a^2}\right]_{\nu_2={\rm const}}
\end{equation}
and
 \begin{equation}
 \label{single- spectrum-short}
 \frac{dw^{(s)}(\nu_1)}{d\nu_1}\propto \exp\left(-\frac{4\nu_1^2}{b^2}\right).
\end{equation}
The coincidence and single-particle spectral widths are defined as FWHM of the curves (\ref{coi-spectrum-short}) and (\ref{single- spectrum-short})
\begin{equation}
 \label{widths-Gaussian-short}
 \Delta\nu_1^{(c)}=2a\sqrt{\ln 2}=\frac{2\,c}{AL}\sqrt{\frac{\ln 2}{\gamma_2}}\,,\quad
 \Delta\nu_1^{(s)}=b\sqrt{\ln 2}=(\ln 2)^{3/4}\sqrt{\frac{2A\omega_0}{\gamma_1\tau B}}\,.
\end{equation}
By comparing these formulas with the earlier derived analytical expressions for spectral widths (\ref{width-short}), we find that all functional dependences in these two sets of formulas are identical. As for numerical coefficients, they can be made identical too if we choose appropriately the fitting parameters $\gamma_1$ and $\gamma_2$ in the Gaussian model substitutes (\ref{subsitution-1}) and (\ref{subsitution-2}) for two factors in the reduced density matrix (\ref{RDM-3}), (\ref{RDM-G-short}). Found from these conditions parameters $\gamma_{1,2}$ are given by
\begin{equation}
 \label{gammas}
 \gamma_1=\sqrt{\ln 2}\approx 0.832555, \quad \gamma_2=\frac{4\,\ln 2}{(5.56)^2}=\frac{\ln 2}{(2.78)^2}\approx 0.0897.
\end{equation}
With these fitting parameters defined, we can write down now all final expressions for the model double-Gaussian wave function, Schmidt number, Schmidt modes and eigenvalues of the reduced density matrix for the case of short pump pulses. So, Equations (\ref{ab-short}) for the parameters $a$ and $b$ become equal to
\begin{equation}
 \label{ab-short-2}
 a=\frac{2.78}{\sqrt{\ln 2}}\frac{c}{AL}=\frac{3.339\,c}{AL}\,,\; b=\sqrt{\frac{2A\omega_0}{\tau B}}\,,
\end{equation}
and the double-Gaussian model (\ref{DG-ab}) of the wave function (\ref{WF}) itself takes the form
\begin{gather}
 \nonumber
 \Psi^{(G)}_{\rm short}= \frac{AL}{\pi c}\sqrt{\frac{\gamma_2}{K_{\rm short}}}\\
 \nonumber
 \times\exp\left[-\frac{\gamma_2A^2L^2(\nu_1+\nu_2)^2}{2c^2}\right]
 \exp\left[-\frac{\gamma_1\tau B(\nu_1-\nu_2)^2}{4A\omega_0\sqrt{\ln 2}}\right]\approx \frac{0.3\,AL}{\pi c\sqrt{K_{\rm short}}}\\
 \label{WF-G-short}
 \times\exp\left[-\frac{0.045\,A^2L^2(\nu_1+\nu_2)^2}{c^2}\right]
 \exp\left[-\frac{0.208\,\tau B(\nu_1-\nu_2)^2}{A\omega_0\sqrt{\ln 2}}\right].
\end{gather}
Comparison of this expression with that of Equation (\ref{WF}) shows that they ar not alike at all. But they do describe a very similar behavior of the wave function $\Psi(\nu_1,\,\nu_2)$. Similarity and differences between the expressions (\ref{WF}) and (\ref{WF-G-short}) are illustrated by two pictures of Figure \ref{Fig2}. In these pictures localization regions of the wave functions (\ref{WF}) and (\ref{WF-G-short}) are shown by thick solid lines. ``Centers of mass$"$ of these lines correspond to maxima (equal to unit) of the narrowest factors in the formulas for these wave functions. For the exact wave function (\ref{WF}) the narrowest factor is the sinc-function, which is maximal when its argument equals zero. This condition yields
\begin{equation}
 \label{c.m.-exact}
 \nu_{1\,{\rm exact}}^{\rm c.m.}(\nu_2)=\nu_2+\frac{A\omega_0}{2B}\left(A-\sqrt{A^2+8BA\nu_2/\omega_0}\right).
\end{equation}
 $\Delta\nu^{(c)}_{\rm shotrt}=5.56\,c/AL$.
\begin{figure}[h]
\centering\includegraphics[width=14cm]{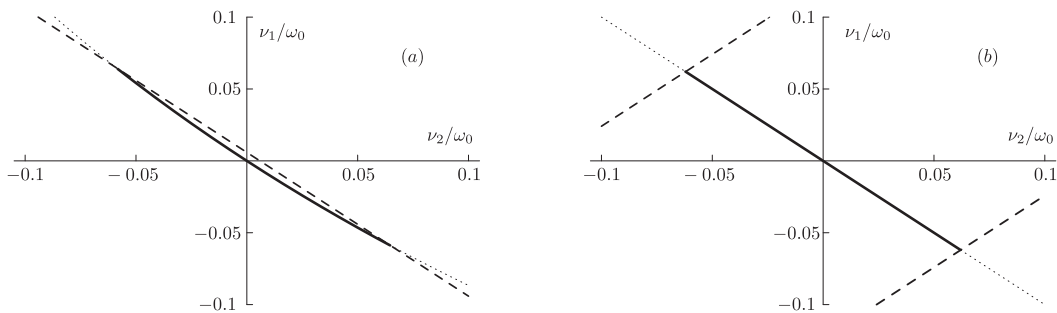}
\caption{{\protect\footnotesize {Thick solid lines determine localization regions of the wave functions $(a)$ (\ref{WF}) and $(b)$ (\ref{WF-G-short}), dashed lines determine  limitations of these regions along the thick solid lines; calculations  for ${\rm LiIO_3}$, pump wavelength $\lambda_0=400\,{\rm nm}$ and pump-pulse duration $\tau = 50\,{\rm fs}$. }}}\label{Fig2}
\end{figure}
For the model double-Gaussian wave function the narrowest part is the first exponential factor on the right-hand side of Equation (\ref{WF-G-short}) and, hence,
\begin{equation}
 \label{c.m.-model}
 \nu_{1\,{\rm Gauss}}^{\rm c.m.}(\nu_2)=-\nu_2.
\end{equation}
Widths of the thick lines in Figure \ref{Fig2} are determined by the coincidence spectral width for short pump pulses (\ref{width-short}), $\Delta\nu_1^{(c)}=5.56\,c/AL$, and, thus, the wave-function localization regions are regions $under$ the thick solid curves in the $(\nu_1,\nu_2)$-map. Note, that in the pictures of Figure \ref{Fig2} widths of the thick solid curves are slightly increased compared to $\Delta\nu_1^{(c)}$ to distinguish clearer these curves from other lines in figures.

Dashed lines in the pictures of Figure \ref{Fig2} determine limitations of the wave-function localization regions in the longitudinal direction imposed by the remaining slowly varying factors in Equations (\ref{WF}) and (\ref{WF-G-short}) [the Gaussian pump spectral function in (\ref{WF}) and the second Gaussian factor on the right-hand side of Equation (\ref{WF-G-short})]. Mathematically these limitations can be found from the condition that the corresponding squared Gaussian function in Equations (\ref{WF}) and (\ref{WF-G-short}) are larger or equal to, e.g.,  1/2, which gives
\begin{equation}
 \label{limitation-exact}
 \nu_{1\,{\rm exact}}^{\rm c.m.}(\nu_2)\leq\nu_{1\,{\rm exact}}^{\rm dashed}(\nu_2),\,
 \nu_{1\,{\rm Gauss}}^{\rm dashed\,-}(\nu_2)\leq\nu_{1\,{\rm Gauss}}^{\rm c.m.}(\nu_2)\leq\nu_{1\,{\rm Gauss}}^{\rm dashed\,+}(\nu_2),
 \end{equation}
where $\nu_{1\,{\rm exact}}^{\rm dashed}(\nu_2)$ and $\nu_{1\,{\rm Gauss}}^{\rm dashed\,\pm}(\nu_2)$ are the functions determining location of the dashed lines, correspondingly, in Figures \ref{Fig2}$(a)$ and \ref{Fig2}$(b)$:
\begin{equation}
 \label{dashed}
  \nu_{1\,{\rm exact}}^{\rm dashed}(\nu_2)=-\nu_2+\frac{2\ln2}{\tau},\,
  \nu_{1\,{\rm Gauss}}^{\rm dashed\,\pm}(\nu_2)=\nu_2\pm(\ln 2)^{1/4}\sqrt{\frac{A\omega_0}{0.208\,B\tau}}.
\end{equation}
These formulas, as well as the pictures of Figure \ref{Fig2}, show clearly that mechanisms, limiting the localization regions of the exact (\ref{WF}) and model (\ref{WF-G-short}) wave functions are absolutely different. But results are seen to be almost identical. Moreover, the only difference in the resulting localization regions in the pictures $(a)$ and $(b)$ of Figure \ref{Fig2} is a very small curvature of the thick solid line in Figure \ref{Fig2}$(a)$ missing in Figure \ref{Fig2}$(b)$. This curvature arises owing to the quadratic term in the argument of the sinc-function in wave function (\ref{WF}). As discussed above, this term is crucially important for determining the single-particle spectrum and limitation of the localization region of the exact wave function (\ref{WF}). In the model function (\ref{WF-G-short}) quadratic terms are missing. Moreover, even if taken into account in some way, such terms are not expected to be important at all. All integral characteristics of the SPDC spectra are identical if calculated with the exact and model double-Gaussian wave functions. In particular, this is true for for the Schmidt number $K$ and single-particle and coincidence spectral widths. As for the Schmidt modes and eigenvalues of the reduced density matrix, it would be very interesting to calculate them numerically to compare with the above derived formulas (\ref{eigenvalues-short}) and (\ref{eigenmodes-short}). We hope to be able making such comparison later.

With the fitting parameters $\gamma_{1,\,2}$ determined by Equation (\ref{gammas}) and the parameters $a$ and $b$ given by Equations (\ref{ab-short-2}), Equation (\ref{K-short-gammas}) for the Schmidt number $K_{\rm short}$ takes the form
\begin{equation}
 \label{K-short}
 K_{\rm short}= \frac{b}{2a}
 =\frac{\sqrt{\pi\ln 2}}{2.78}\,\frac{A^{3/2}}{\sqrt{B}}\frac{L}{\sqrt{\lambda_0c\tau}}=0.5308\,
 \frac{A^{3/2}}{\sqrt{B}}\frac{L}{\sqrt{\lambda_0c\tau}}\,.
\end{equation}
This expression coincides exactly with that of the work \cite{Spectral}  for the parameter $R$ in the regime of short pulses.

The reduced density matrix (\ref{RDM-G-short}), after all substitutions, takes its final form
\begin{gather}
 \nonumber
 \rho_{r,\, {\rm short}}(\nu_1,\,\nu_1^{\,\prime}) =\sqrt{\frac{2B\tau}{\pi A\omega_0}}\\
 \label{RDM-G-short-final}
\times\exp\left[-\frac{B\tau}{2A\omega_0}(\nu_1+\nu_1^{\,\prime})^2\right]\, \exp\left[-\frac{A^2L^2\ln 2}{(5.56\,c)^2}(\nu_1-\nu_1^{\,\prime})^2\right].
\end{gather}
For the parameter $\mu$, determining eigenvalues of the reduced matrix, we get formally the same expression as in the case of long pump pulses, $\mu\approx -1+2\frac{a}{b}$, though with different constants $a$ and $b$, which gives finally
\begin{equation}
 \label{mu-short}
 \mu\approx -1+\frac{1}{K_{\rm short}}=-1+\frac{2.78}{\sqrt{\pi\ln 2}}\,\frac{\sqrt{B}}{A^{3/2}}\frac{\sqrt{\lambda_0c\tau}}{L},
\end{equation}
and eigenvalues of the reduced density matrix are given by
\begin{equation}
 \label{eigenvalues-short}
 \lambda_n\approx \frac{2}{K_{\rm short}}\left(1-\frac{2}{K_{\rm short}}\right)^n
\end{equation}
with $K_{\rm short}$ determined by Equation (\ref{K-short}).

The scaling factor $\alpha$ of the Schmidt modes (\ref{eigenmodes}) in the case of short pump pulses has the form
\begin{equation}
 \label{alpha-short}
 \alpha_{\rm short}=\frac{1}{a\sqrt{K_{\rm short}}}=\frac{AL}{c}\frac{\sqrt{\ln 2}}{2.78\,\sqrt{K_{\rm short}}}
 =0.299\,\frac{AL}{c\sqrt{K_{\rm short}}},
\end{equation}
and the Schmidt modes are given by
\begin{gather}
 \nonumber
 \psi_n(\nu_{1,\,2})=\sqrt{\frac{AL}{c}}
 \frac{(\ln 2)^{1/4}}{\sqrt{2.78}\,(K_{\rm short})^{1/4}}
 \,\varphi_n\left(\frac{\nu_{1,\,2}AL}{c}\frac{\sqrt{\ln 2}}{2.78\,\sqrt{K_{\rm short}}}\right)\\
 \label{eigenmodes-short}
 \approx
 0.547\,\sqrt{\frac{AL}{c\sqrt{K_{\rm short}}}}\;\varphi_n\left(0.299\,\frac{\nu_{1,\,2}AL}{c\sqrt{K_{\rm short}}}\right).
\end{gather}

\section{Generalizations}

By comparing Equations (\ref{mu-long}) and (\ref{eigenvalues-long}) with (\ref{mu-short}) and (\ref{eigenvalues-short}), we find that they look identical with the only substitution: $K_{\rm long}\rightleftharpoons K_{\rm short}$. So, we can suggest the following simplest interpolation rule for a transition between the regions of short and long pulses consisting in the assumption that even in the intermediate region ($\eta\sim 1$) the dependence of $\mu$ and $\lambda_n$ of the Schmidt number $K$ remains the same as in the asymptotic regions $\eta\ll 1$ and $\eta\gg 1$:
\begin{gather}
 \nonumber
 \mu(\tau)=-1+\frac{1}{K(\tau)},\\
 \label{mu-lambda-general}
 \lambda_n(\tau)= \frac{2}{K(\tau)}\left(1-\frac{2}{K(\tau)}\right)^n
 \approx\frac{2}{K(\tau)}\,\exp\left(-\frac{2\,n}{K(\tau)}\right) .
\end{gather}
Of course,these relations are proved rigorously only in the limits of short ($\eta\ll 1$) and long ($\eta\gg 1$) pump pulses. But as the Schmidt number $K(\tau)$ is large also in the case of intermediate pulse durations, the assumption that Equations (\ref{mu-lambda-general}) remain valid also in the case $\eta\sim 1$ seems rather natural. Note however that the function $K(\tau)$ in the region of intermediate pulse durations either has to be taken from numerical calculations \cite{Mauerer} or in its turn has to be defined by means of a more or less reasonable interpolation. The simplest interpolation $K(\tau)=\sqrt{K_{\rm short}^2(\tau)+K_{\rm long}^2(\tau)}$ was suggested and used in our earlier work \cite{Spectral}. Below we discuss somewhat more elaborate ways of interpolating $K(\tau)$ into the region $\eta\sim 1$.

Following the same logic, let us make now a rather important assumption that the double-Gaussian modeling of the wave function (\ref{WF}) has sense not only in the asymptotic cases of short and long pump pulses but also in the case of pulses of intermediate duration, when $\eta\sim 1$. In other words, we assume that for any values of the control parameter $\eta$, small, intermediate, and large, the biphoton wave function (\ref{WF}) can be satisfactorily modeled by
\begin{equation}
 \label{general model}
 \Psi(\nu_1,\nu_2;\,\tau)=\sqrt{\frac{2}{\pi a(\tau) b(\tau)}}\,\exp\left[-\frac{(\nu_1+\nu_2)^2}{2a(\tau)^2}\right]\, \exp\left[-\frac{(\nu_1-\nu_2)^2}{2b(\tau)^2}\right],
\end{equation}
where the widths $a(\tau)$ and $b(\tau)$ are determined by Equations (\ref{ab-long}) and (\ref{ab-short-2}) in the cases of long and short pump pulses and have to be defined yet in the case of intermediately long pulses. By comparing expressions in Equations (\ref{ab-long}) and (\ref{ab-short-2}) we find easily that they are related to each other by the following simple formulas
\begin{gather}
 \nonumber
 \frac{a_{\rm short}}{a_{\rm long}}=\frac{1.39}{2\ln 2}\,\eta=1.00267\,\eta\approx\eta\quad{\rm and}\\
 \label{ab-short-long}
 \frac{b_{\rm long}}{b_{\rm short}}=\sqrt{\frac{c\tau}{2\times 0.249\,LA}}=1.002\sqrt{\eta}\approx\sqrt{\eta}.
\end{gather}
By using the idea of interpolation we assume that in a general case we can define the functions $a(\tau)$ and $a(\tau)$ as being given by
\begin{equation}
 \label{Interp-ab}
 a(\tau)=a_{\rm short}f_a(\eta)\quad{\rm and}\quad b(\tau)=b_{\rm long}f_b(\eta),
\end{equation}
where the smoothing functions obey the conditions
\begin{equation}
 \label{smoothing-limiting}
 \left.f_a(\eta)\right|_{\eta\ll 1}=1,\;\left.f_a(\eta)\right|_{\eta\gg 1}=\frac{1}{\eta};\quad
 \left.f_b(\eta)\right|_{\eta\ll 1}=\frac{1}{\sqrt{\eta}},\;\left.f_b(\eta)\right|_{\eta\gg 1}=1.
\end{equation}
Of course, there are infinitely many ways of choosing the smoothing functions $f_a(\eta)$ and $f_b(\eta)$. To restrict somehow this manifold of possibilities, let us take these functions in the form
\begin{equation}
 \label{smooth-func}
 f_a(\eta)=\left(1+\eta^s\right)^{-1/s}\quad{\rm and}\quad f_b(\eta)=\frac{\left(1+\eta^s\right)^{1/2s}}{\sqrt{\eta}},
\end{equation}
where $s>0$ is a free parameter. With these functions $f_a(\eta)$ and $f_b(\eta)$ Equations (\ref{Interp-ab}) are reduced to
\begin{gather}
 \nonumber
 a(\tau)=a_{\rm short}\left(1+\eta^s\right)^{-1/s};\;
 \frac{a(\tau)}{\omega_0}=\frac{1.39}{\pi A\sqrt{\ln 2}}\,\frac{\lambda_0}{L}\,\left(1+\eta^s\right)^{-1/s},\\
 \label{Interp-ab}
 b(\tau)=b_{\rm long}\frac{\left(1+\eta^s\right)^{1/2s}}{\sqrt{\eta}};\;
 \frac{b(\tau)}{\omega_0}=\sqrt{\frac{\lambda_0}{2\pi\, 0.249\,BL}}\frac{\left(1+\eta^s\right)^{1/2s}}{\sqrt{\eta}}\,.
\end{gather}
The dependences $a(\tau)$ and $b(\tau)$ are characterized by two curves in Figure \ref{Fig3}. The dashed parts of these curves characterize smooth transitions from the short-pulse to long-pulse regions.
\begin{figure}[h]
\centering\includegraphics[width=14cm]{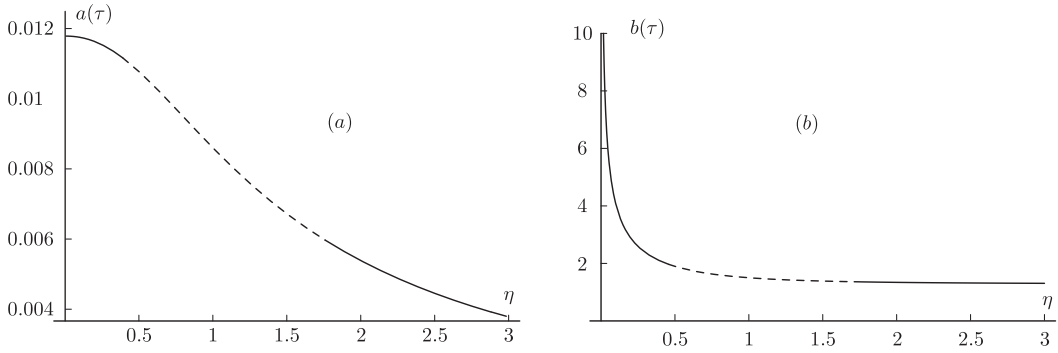}
\caption{{\protect\footnotesize {Parameters $a(\tau)$ and $b(\tau)$ (\ref{Interp-ab}) of the general double-Gaussian spectral wave function (\ref{general model}), calculations for LiIO$_3$ crystal, $L=0.5\,{\rm cm}$; dashed lines are interpolation with the help of the functions (\ref{smooth-func}) with $s=2.21$.  }}}\label{Fig3}
\end{figure}
One of the main features of the widths $a(\tau)$ and $b(\tau)$ seen clearly in Figure \ref{Fig3} is that at all values of the pump-pulse duration $\tau$, or of the control parameter $\eta$, always $b(\tau)\gg a(\tau)$. This feature of the model wave function (\ref{general model}) differs significantly from that of the exact wave function (\ref{WF}). In the latter the pump spectral function and sinc-function change their roles at $\eta\sim 1$: in the region $\eta\ll 1$ the sinc-function is much narrower and in the region $\eta\gg 1$ much wider than the pump spectral function. In contrast to this, in the case of the model double-Gaussian wave function, the first factor on the right-hand side of Equation (\ref{general model}) is much narrower than the second one at all values of the pump-pulse duration $\tau$. This simplifies, for example, the definition of the coincidence and single-particle spectral widths. In terms of $a(\tau)$ and $b(\tau)$ they are given by
\begin{equation}
 \label{coi-sibngle-model}
 \Delta\nu_1^{(c)}(\tau)=2a(\tau)\,\sqrt{\ln 2},\quad\Delta\nu_1^{(s)}(\tau)=b(\tau)\,\sqrt{\ln 2}\,,
\end{equation}
for all $\tau$.

With the definitions of Equations (\ref{Interp-ab}) we get the following generalized expression for the Schmidt number
\begin{gather}
  \label{K(tau)}
  K(\tau)=\frac{b(\tau)}{2a(\tau)}=\frac{\sqrt{\pi\ln 2}}{2.78\sqrt{2\times 0.249}}\,A\,\sqrt{\frac{L}{B\lambda_0}}
 \,\frac{[1+\eta^s(\tau)]^{3/2s}}{\sqrt{\eta(\tau)}}.
\end{gather}
The dependence $K(\tau)$ (or $K(\eta)$) is similar to that characterized by the curves in Figure \ref{Fig1}. The minimum of the function $K(\tau)$ is achieved at $\eta_0=2^{-1/s}$ and
\begin{gather}
  \label{K(tau)min}
  \left.K(\tau)\right|_{\min}=\frac{\sqrt{\pi\ln 2}}{2.78\sqrt{2\times 0.249}}\,A\,\sqrt{\frac{L}{B\lambda_0}}
 \,\frac{3^{3/2s}}{2^{1/s}}.
\end{gather}
The case $s=3$ corresponds to the earlier used interpolation \cite{Spectral}. If we want to increase $\left.K(\tau)\right|_{\min}$  by 12$\%$ to make the curve $K(\tau)$ closer to the numerically calculated one \cite{Mauerer}, we have to take $s=2.21$ in our formulas (\ref{smooth-func})-(\ref{K(tau)min}). This gives $\left.K(\tau)\right|_{\min}=83$ (instead of $\left.K(\tau)\right|_{\min}=73$ of the work \cite{Spectral}), and the curve $K(\eta)$ appears in this case practically indistinguishable from the numerically calculated one. But of course, the main novelty of the described here method of interpolation is not only in a possibility of getting this 12$\%$ improvement. The main message is in the assumption about a possibility of modeling the wave function (\ref{WF}) at arbitrary values of the pump-pulse duration $\tau$ by the double-Gaussian wave function (\ref{general model}) with the parameters $a(\tau)$ and $b(\tau)$ given by Equations (\ref{Interp-ab}). This assumption is proved to be correct in the cases of long and short pulses. At the ``integral level$"$, correctness of the suggested model (\ref{general model}) is also proved for all values of $\tau$ because the Schmidt number $K(\tau)$ is found to be practically identical in the exact numerical calculations \cite{Mauerer} and in the given above analytical derivation for the model double-Gaussian wave function (\ref{general model}) with $s=2.21$. Localization regions of the exact (\ref{WF}) and model (\ref{general model}) wave functions in the case of intermediate pump-pulse durations are also very close to each other and, qualitatively, similar to that shown in Figure \ref{Fig2}$(b)$. A more detailed comparison of the exact (\ref{WF}) and model (\ref{general model}) wave functions in the case of intermediately long pump pulses can involve comparison of eigenvalues of the reduced density matrix and of a structure of the Schmidt modes. But for the exact wave function (\ref{WF}) at $\eta\sim 1$ such numerical data are not obtained yet.

Returning to the generalized formula (\ref{mu-lambda-general}) for the eigenvalues $\lambda_n(\tau)$ of the reduced density matrix, as said above, we assume that this expression is valid for not only for small and large $\eta$ but also also in all the region between these two asymptotic limits. In accordance with  the Schmidt theorem, the weights with which the Schmidt modes are represented in the expansion (\ref{Schmidt theorem}) are given by $\left|\sqrt{\lambda_n(\tau)}\right|$. As an example, let us consider the case $\eta=1$, when $K(\tau)=87$. In this case the values of $\left|\sqrt{\lambda_n(\tau)}\right|_{\eta=1}$ are located along the line shown in Figure \ref{Fig4}. At $n=K=87$, we get $\left|\sqrt{\lambda_{87}(\tau)}\right|_{\eta=1}=0.055$, which as high as $37\%$ of the maximal value of $\left|\sqrt{\lambda_n}\right|$, $\left|\sqrt{\lambda_1}\right|=0.15$. This results show that falling of $\left|\sqrt{\lambda_n}\right|$ with a growing $n$ is rather slow, and in reality there is a rather large interval of $n>K$, giving not too small contribution into the expansion of the wave function in the sum of the Schmidt-mode products (\ref{Schmidt theorem}).
\begin{figure}[h]
\centering\includegraphics[width=7cm]{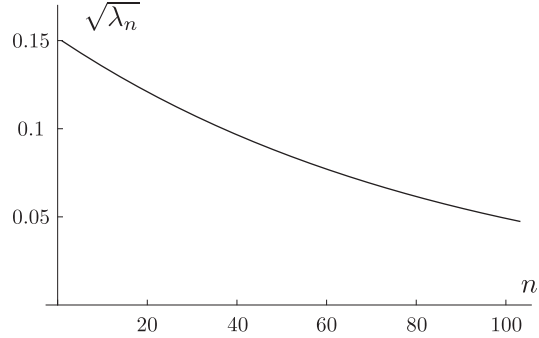}
\caption{{\protect\footnotesize {Square roots of the eigenvalues of the reduced density matrix (\ref{mu-lambda-general}) at $\eta=1$ and $K(\tau)$ given by Equation (\ref{K(tau)}).  }}}\label{Fig4}
\end{figure}

 The Schmidt modes of the model double-Gaussian wave function (\ref{general model})  have the same form as in the asymptotic cases of long (\ref{eigenmodes-long}) and short (\ref{eigenmodes-short}) pump pulses, but now with the generalized parameters $a(\tau)$ and $b(\tau)$ (\ref{Interp-ab}):
\begin{gather}
 \label{Schmidt-modes-generalized}
 \psi_n(\nu_{1,\,2};\,\tau)=\frac{1}{\sqrt{\alpha(\tau)}}\,\varphi_n\left[\omega_0\alpha(\tau)\frac{\nu_{1,\,2}}{\omega_0}\right],
\end{gather}
where $\omega_0\alpha(\tau)$ is the dimensionless generalized scaling factor characterizing the dependence of $\psi_n$ on the dimensionless variables $\nu_{1,\,2}/\omega_0$:
\begin{gather}
 \nonumber
 \label{alpha-generalized}
 \omega_0\alpha(\tau)=\omega_0\sqrt{\frac{2}{a(\tau)b(\tau)}}
   =2.17\,\sqrt{A}B^{1/4}\left(\frac{L}{\lambda_0}\right)^{3/4}\,\eta^{1/4}(1+\eta^s)^{1/4s}.
\end{gather}
As a function of the control parameter $\eta$, the dimensionless scaling factor of the Schmidt modes is plotted in Figure \ref{Fig5}.
\begin{figure}[h]
\centering\includegraphics[width=7cm]{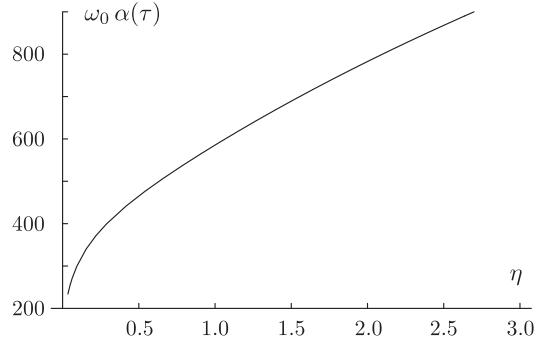}
\caption{{\protect\footnotesize {Dimensionless scaling factor $\omega_0\alpha(\tau)$ (\ref{alpha-generalized}) vs. the control parameter $\eta$, for LiIO$_3$ crystal, $L=0.5\,{\rm cm}$, $\lambda_0=400\,{\rm nm}$, $s=2.21$.  }}}\label{Fig5}
\end{figure}
The dimensionless scaling factor is large practically at all values of $\eta$. With a growing $\eta$ the scaling factor monotonously grows, which means that localization regions of all Schmidt modes shrink. At $\eta=1$, Eq. (\ref{alpha-generalized}) gives $\omega_0\alpha(\tau)\approx 585.5$.

The spectral width of an $n^{\rm th}$ Schmidt mode (\ref{Schmidt-modes-generalized}) can be estimated as
\begin{equation}
 \label{SM-spectrum}
 \delta\nu_n(\tau)\sim\frac{\sqrt{n}}{\alpha(\tau)}
 =\sqrt{\frac{n\,a(\tau)b(\tau)}{2}}.
\end{equation}
This width grows with the increasing mode number $n$ as $\sqrt{n}$. For $n=K(\tau)$ Equation (\ref{SM-spectrum}) gives $\delta\nu_K\sim b(\tau)/2=\Delta\nu^{(s)}/2\sqrt{\ln 2}=0.6\, \Delta\nu^{(s)}$; $\delta\nu_n(\tau)$ reaches $\Delta\nu^{(s)}$ at $n\approx 2.77\,K(\tau)$. It's true, however,  that the contribution of such high-number Schmidt modes to the expansion (\ref{Schmidt theorem}) is very small: for the same parameters as used above for all estimates, we get $\sqrt{\lambda_{2.77\,K}}\sim 6\%$ of $\sqrt{\lambda_1}$ (to be compared with $\sqrt{\lambda_{K}}\sim 37\%$ of $\sqrt{\lambda_1}$). This means that practically for all important Schmidt modes their spectral width is smaller than the single-particle width of the biphoton spectrum as a whole.

As an example, some spectral Schmidt modes are shown in Figure \ref{Fig6} for the same parameters as in Figure \ref{Fig5} and $\eta=1$.
\begin{figure}[h]
\centering\includegraphics[width=14cm]{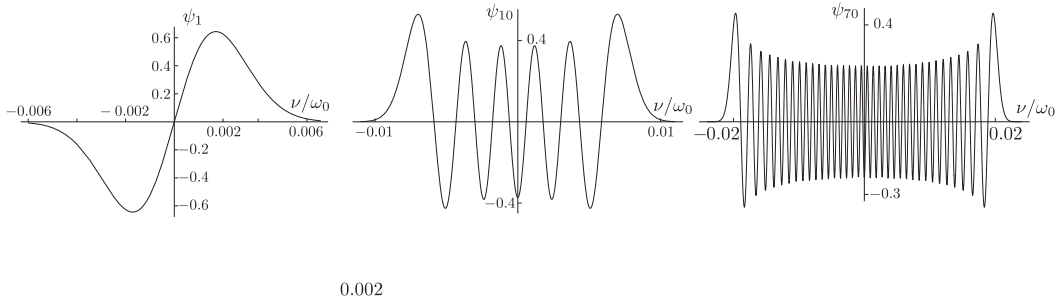}
\caption{{\protect\footnotesize {Schmidt modes $\psi_1$, $\psi_{10}$ and $\psi_{70}$.  }}}\label{Fig6}
\end{figure}

\section{Conclusion}

Summarizing, we found double-Gaussian models of the SPDC type-I wave function (\ref{WF}) in both regions of short and long pump pulses. In the case  of short pump pulses even a possibility of Gaussian modeling was not evident in advance. Its derivation required a transition to the reduced density matrix where Gaussian modeling appears to be much more natural and simple than in the case of the wave function. Then, after finding a double-Gaussian model for the reduced density matrix and using general relations between its parameters and those of the wave function, we were able to find the double-Gaussian model (\ref{WF-G-short}) of the original wave function (\ref{WF}) in the case of short pump pulses. Though the forms of these too functions are significantly different, the double-Gaussian model reproduces a structure of the exact wave function rather well and almost in all details, which is seen clearly in two pictures of Figure \ref{Fig2}. Having the double-Gaussian models of the exact wave function in the regions of short and long pulses we found also in these two cases the Schmidt modes (\ref{eigenmodes-long}) and (\ref{eigenmodes-short}). Finally, by analyzing features of all parameters in the asymptotic cases of long and short pump pulses and using the most appropriate interpolation to the region of intermediately long pump pulses, we found a possibility of suggesting the model of a double-Gaussian wave function [Equations (\ref{general model}) and (\ref{Interp-ab})], which we assume to be valid for arbitrary values of the pump-pulse duration $\tau$. For this model wave function we found also the Schmidt number $K(\tau)$ (\ref{K(tau)}), Schmidt modes [Equations (\ref{Schmidt-modes-generalized}) and (\ref{alpha-generalized})], and eigenvalues of the reduced density matrix (\ref{mu-lambda-general}). The Schmidt number  found in such a way agrees perfectly well with the results of its numerical calculation \cite{Mauerer}. We assume that the functions $\psi_n(\nu_{1,\,2})$ of Equation (\ref{Schmidt-modes-generalized}) represent equally well the spectral Schmidt modes in all range of the pump-pulse durations, short, long and intermediate. Comparison with exact numerical calculations of the Schmidt modes is expected to be interesting and fruitful. Also we hope that the derived results can be tested experimentally. For example, an interesting scheme of an experiment can involve splitting of the biphoton beam for two channels by a nonselective beam-splitter, installing in one of two channels a spectral mask gating through it only a single spectral mode and measuring the coincidence spectrum in the second channel. According to the Schmidt theorem the coincidence spectrum measured in such a way is identical to that of the mode gated through in the first channel, and this coincidence of two spectra deserves its experimental verification. We hope also that in future the data about spectral Schmidt modes can be used for practical applications in the problems of quantum information and quantum cryptography.

\end{document}